\documentclass[onecolumn,preprintnumbers,amsmath,amssymb]{revtex4}
\usepackage{epsfig,graphicx}
\usepackage{dcolumn}
\usepackage{bm}
\begin{document}
\draft
\title{Electronic Properties of TiO$_2$ Nanotubes}

\vspace{1cm}
\author{V.V. Ivanovskaya*, A.N. Enyashin, N.I. Medvedeva and A.L. Ivanovskii\footnote[1]{For correspondence: viktoria@ihim.uran.ru}}
\address{Institute of Solid State Chemistry,\\ Ural Branch of the Russian Academy of
Sciences,\\Ekaterinburg, 620219, Russia}

\begin{abstract}

New quasi-one-dimensional (1D) titania nanostructures -
single-walled nanotubes formed by rolling [101] planes of TiO$_2$
(anatase phase) are modeled and their electronic properties and
bond orders indices are studied using the tight-binding band
theory. We show that all zigzag (n,0)- and armchair (n,n)-like
nanotubes are uniformly semiconducting, and the band gap trends
to vanish as the tube diameters decrease. It was established that
the zigzag (n,0) nanotubes configurations are more likely to form
when the diameters are larger 1 nm. The Ti-O covalent bonds were
found to be the strongest interactions in TiO$_2$ tubes, whereas
Ti-Ti bonds proved to be much weaker.

\end{abstract}
\maketitle


One-dimensional (1D) nanostructured materials (nanotubes) have
been the focus of extensive research due to their technological
importance. Alongside there are the well-known classes of
sp$^2$-bonded planar structures (graphite, BN, BC$_3$, BC$_2$N
etc) which can be rolled into cylinders (reviews
\cite{dresselhaus,saito,moriarty,knupfer,ajayan,ivanovskii}),
some efforts are reported also in synthesis and studying of
d-metal (M) containing inorganic nanotubes (NT). Nanotubes of
layered dichalcogenides MX$_2$ (M = Mo, W, Ta, Nb etc.; X = S, Se)
have been produced
\cite{rothschild,tenne,nath,rosentsveig,tenne2}, and some
microscopic models of nanotubular structures formed from layered
diborides MB$_2$ (M = Sc, Ti, Zr) have been proposed recently
\cite{chernozatonskii,ivanovskaya}.

On the other hand, much interest was aroused recently in 1D
nano-structures (nanowires and nanotubes) based on non-layered
semiconducting d-metal oxides such as V$_2$O$_5$
\cite{krumeich,krumeich2,muller,muhr,niederberger,pillai},
Co$_3$O$_4$ \cite{shi,shi2}, MnO$_2$ \cite{lakshmi}, WO$_3$
\cite{lakshmi} etc. Among them 1D nanostructured titania materials
have considerable scientific and technological significance due to
their chemical inertness, endurance, strong oxidizing power,
non-toxicity and lower production cost. Important progress was
achieved in synthesis of polycrystalline TiO$_2$ nanowires and
nanotubes using sol-gel template approaches
\cite{imai,shimizu,kasuga,imai2,seo,miao}.

Recently, the first synthesis of single-crystalline TiO$_2$ NTs
has been performed \cite{liu} using unconstrained solution growth
method by hydrolyzing TiF$_4$ under acidic condition without
resort to solid templates. Those titania multi-walled nanotubes
are constructed with concentric stacking of [101] planes of the
anatase polymorph. The diameters (D) of the inner and outer walls
are within the range 2,5-5,0 and 20-40 nm, respectively. However,
the electronic properties of the titania nanotubes have not been
studied up till now.

In this Letter we report the results of our band structure
calculations of titania nanotubes constructed from single planes
of TiO$_2$ (anatase). Their electronic properties and bond indices
are analyzed as a function of tubes diameters (D) and
configuration ({\it armchair- and zigzag-}like).

The anatase structure (space group {\it I41/amd}) is build up by
TiO$_6$ units. The lattice constants of TiO$_2$ equal to {\it a}
= 0,3782, {\it c} = 0,9502 nm and the internal parameter {\it u =
d$_a$/c}, where d$_a$ is the apical Ti-O bond length (0,1979 nm)
\cite{burdett}.

The structure of the [101] layer (nominal stoichiometry TiO$_2$)
is shown in Fig. 1. Similar to graphene sheets
\cite{dresselhaus,saito,moriarty,knupfer,ajayan,ivanovskii,rothschild}),
sandwich layers (S-M-S)
\cite{tenne,nath,rosentsveig,tenne2,seifert,seifert2}) or
metal-boron bilayers \cite{chernozatonskii,ivanovskaya}) in
carbon, MS$_2$ or MB$_2$-based NTs, respectively, these layers
can be mapped onto the surface of a cylinder forming
"triple-wall" (O-Ti-O) tubes, Fig. 2. As in single-walled carbon
NTs \cite{saito,moriarty,knupfer,ajayan}), depending on the
rolling direction {\bf c} in the 2D lattice {\it {\bf c} = n{\bf
a$_1$}} + {\it m{\bf a$_2$}} ({\it {\bf a$_1$}}, {\it {\bf
a$_2$}} are primitive vectors for the honeycomb lattice), three
groups of TiO$_2$ 1D-structures can be constructed: {\it
armchair}({\it n},{\it n})-, {it zigzag} ({\it n},0)-like and
{\it chiral} ({\it n},{\it m}) nanotubes.

Our calculations were performed for ({\it n},0) and ({\it n},{\it
n}) titania NTs as a function {\it n} in ranges from (8,0) to
(15,0) and from (4,4) to (15,15), which correspond to the
intervals of the inner diameters (D$^i$$^n$) 0,40 - 1,08 and 0,28
- 2,06 nm, respectively, see Table 1. The largest diameters of our
model tubes are comparable with the smallest experimentally
observed TiO$_2$ tubes diameters \cite{liu}.

The tight binding band structure method within extended Huckel
theory (EHT) approximation \cite{hoffmann} was employed. Besides
the electronic band structure, this approach allows to
investigate the chemical bonding based on the Mulliken analysis
scheme. The total densities of states (DOS), crystal orbital
overlap populations (COOP), and total band energies of the
nanotubes (E$_t$$_o$$_t$) were obtained. The calculated DOS of
some ({\it n},{\it n}) and ({\it n},0)-like TiO$_2$ NTs (Fig. 3)
are similar for all titania tubes and also agree with DOS of the
crystalline anatase \cite{asahi}. The valence band is composed of
two electronic bands separated by a forbidden gap. The lowermost
quasi-core band located $\sim$17-16 eV below the Fermi level (not
shown in Fig. 3) consists mainly of O2s states. The near-Fermi
Ti3d-O2p-band is fully occupied, and the lower part of the
conduction band is made up predominantly of Ti3d states. The
lower energy part of the hybrid band is formed by mixed
Ti3d-O2p($\sigma$)-states, whereas the upper edge - by
non-bonding O2p($\pi$) states. Their distribution depends on the
tubes' geometry and diameter and determines the variation in the
band gap (BG) width. For the considered NTs with the maximal
diameter (minimal curvature), the BG values are $\sim$3,34 ({\it
armchair}-like (15,15)NT, D$^i$$^n$ = 2,06 nm) and $\sim$3,01 eV
({\it zigzag}-like (15,0)NT, D$^i$$^n$ = 1,08 nm) as compared
with the calculated \cite{asahi} value of $\sim$2,0 eV (direct
transition at $\Gamma$) and experimentally determined \cite{tang}
band gap width of crystalline anatase ($\sim$3,4 eV). The BG
decreases with diminishing NT diameters (Fig. 4a), and for {\it
armchair}-like NTs it being still higher than those for {\it
zigzag}-like NTs. It is of interest that a similar dependence in
BG as a function of the tube diameter and configuration was
reported for semiconducting NTs based on layered MoS$_2$ and
WS$_2$ \cite{seifert,seifert2}.

Figure 4b shows the calculated values of E$_t$$_o$$_t$ (per
TiO$_2$ unit) versus titania NT diameters. The E$_t$$_o$$_t$
dependence follows a$\sim$1/D$^2$ behavior indicative of a
decrease in TiO$_2$ tubes stability with decreasing D. Analogous
dependencies of strain energy (the difference between the
energies of the plane atomic layer and the corresponding NT
characterizes the chemical stability of tubular structures) are
known for carbon and the majority of non-carbon tubes
\cite{dresselhaus,saito,moriarty,knupfer,ajayan,ivanovskii,rothschild,tenne,nath,chernozatonskii,seifert,seifert2}.
It is worth noting that our results show that for small diameters
(D$^i$$^n$ $<$ 1 nm) armchair-like configurations are more
stable, whereas for large diameters zigzag-like configurations of
titania NTs have greater stability, Fig. 4b.

The formation of the covalent bonds in a tube can be readily seen
from the COOP values (Table 1). For all tubes, (I) the main bonds
are Ti-O interactions; (II) the occupation of the Ti-Ti bonds is
by an order of magnitude smaller; (III) O-O bonding are absent
(COOPs $<$ 0); and (IV) there is a sharp anisotropy of separate
Ti-O bonds depending both on their orientation relative to the
TiO$_2$ tube axis and between the bonds of titanium and oxygen
atoms belonging to inner (O$^i$$^n$) or outer (O$^o$$^u$$^t$)
oxygen cylinders. For {\it zigzag}-like tubes, the main bonds are
Ti-O$^o$$^u$$^t$. As D increases, their anisotropy decreases. For
{\it armchair}-like tubes, the COOPs values of Ti-O$^i$$^n$ and
Ti-O$^o$$^u$$^t$ are comparable.

It should be noted that the total energy versus diameter
dependence is almost saturated at D$>$1,2 nm and the asymptotic
value of E$_t$$_o$$_t$  is slightly depends from diameter up to
2,4 nm. The weak dependence of BG from diameter for large values
of D is also observed (Fig 4a).

In summary, single-walled titania nanotubes with nominal
stoichiometry TiO$_2$ were modeled and their electronic properties
and bonding indices have been investigated using the
tight-binding band theory. We show that both {\it zigzag}- and
{\it armchair}-like nanotubes are semiconducting, and the band gap
trends to vanish for very small NT diameters. {\it Zigzag} titania
nanotubes were found to form more readily for diameters
comparable with those observed experimentally \cite{liu}. It was
established that Ti-O covalent bonds are the strongest in TiO$_2$
tubes, whereas Ti-Ti bonds are much weaker.

There are numerous issues of interest for future studies. First
of all, an important problem is the effect of "interlayer"
interactions on the properties of milti-walled TiO2 tubes, as
well as simulation of the types of polyhedra, which close the
open ends of the tubes. A probable factor of titania NT
electronic properties variation is "polygonization" of their
walls (flattened segments of [101] planes in the tube walls were
obseved experimentally \cite{liu}), which calls for
investigations of radially deformed tubes.

\newpage

\begin{table}
\begin{center}
\caption{The number of atoms in unit cells, diameters (D, nm) and
indices of intra-atomic bonds (COOPs, e) for TiO$_2$ nanotubes.}

\begin{tabular}{|c|c|c|c|c|c|c|c|c|c|}
\hline &&&& \multicolumn{6}{c|}{COOPs**}
\\\cline{5-10}
Tubes&Cell size& D$^i$$^n$& D$^o$$^u$$^t$& Ti-O$^i$$^n$$_1$&
Ti-O$^i$$^n$$_2$&
Ti-O$^o$$^u$$^t$$_1$& Ti-O$^o$$^u$$^t$$_2$& Ti-Ti$_1$& Ti-Ti$_2$\\

\hline (8,0)&48&0,422&0,896&0&0&0,499&0,331&0,041&0,028\\
\hline (9,0)&54&0,518&0,992&0&0&0,488&0,349&0,021&0,005\\
\hline (10,0)&60&0,614&1,082&0&0&0,477&0,351&0,018&0,001\\
\hline (11,0)&66&0,708&1,174&0&0&0,470&0,353&0,018&0\\
\hline (12,0)&72&0,800&1,266&0&0&0,464&0,354&0,017&0\\
\hline (13,0)&78&0,894&1,356&0&0&0,458&0,355&0,017&0\\
\hline (14,0)&84&0,986&1,448&0&0&0,453&0,355&0,017&0\\
\hline (15,0)&90&1,076&1,538&0&0&0,449&0,356&0,017&0\\
\hline (4,4)&24&0,286&0,782&0,117&0,302&0,241&0,302&0,004&0,036\\
\hline (5,5)&30&0,464&0,946&0,181&0,279&0,235&0,308&0&0,031\\
\hline (6,6)&36&0,632&1,108&0,218&0,285&0,229&0,310&0&0,029\\
\hline (7,7)&42&0,798&1,268&0,242&0,292&0,224&0,312&0&0,029\\
\hline (8,8)&48&0,960&1,426&0,258&0,298&0,219&0,313&0&0,029\\
\hline (9,9)&54&1,120&1,584&0,270&0,302&0,216&0,314&0&0,029\\
\hline (10,10)&60&1,278&1,742&0,279&0,305&0,213&0,315&0&0,030\\
\hline (11,11)&66&1,436&1,900&0,287&0,307&0,210&0,315&0&0,030\\
\hline (12,12)&72&1,594&2,056&0,293&0,309&0,208&0,316&0&0,030\\
\hline (13,13)&78&1,750&2,212&0,298&0,310&0,206&0,316&0&0,030\\
\hline (14,14)&84&1,906&2,368&0,303&0,311&0,204&0,316&0&0,030\\
\hline (15,15)&90&2,064&2,524&0,306&0,312&0,202&0,316&0&0,030\\
\hline
\end{tabular}

 * D$^i$$^n$, D$^o$$^u$$^t$ - diameters of "inner" and "outer" cylinders
made up of oxygen atoms.

** Indices 1, 2 correspond to paired bonds with a different
orientation along the tube axis (see Fig. 1); Ti-O$^i$$^n$ and
Ti-O$^o$$^u$$^t$ are the bonds of Ti with atoms of the "inner"
and "outer" oxygen cylinders.
\end{center}
\end{table}

\begin{figure}
\includegraphics[width=0.65\textwidth]{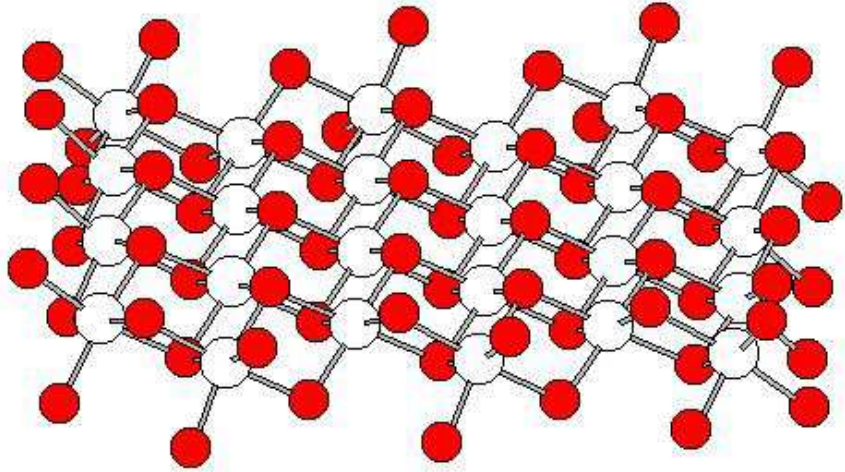}
\caption{Top view on the structure of [101] layer TiO$_2$
(anatase).}
\end{figure}

\begin{figure}
\includegraphics[width=0.65\textwidth]{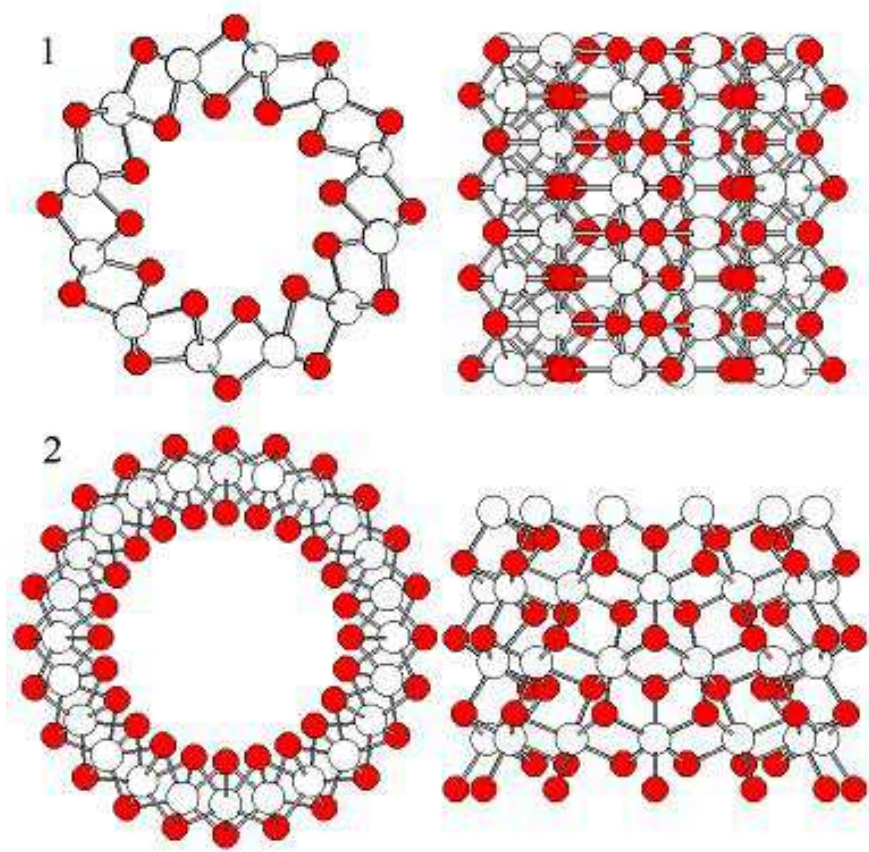}
\caption{The structure of 1-{\it armchair} ({\it 6},{\it 6}) and
2-{\it zigzag} ({\it 12},{\it 0}) TiO$_2$ NTs. Side views and
views along the tube axis are shown.}
\end{figure}

\begin{figure}
\includegraphics[width=0.65\textwidth]{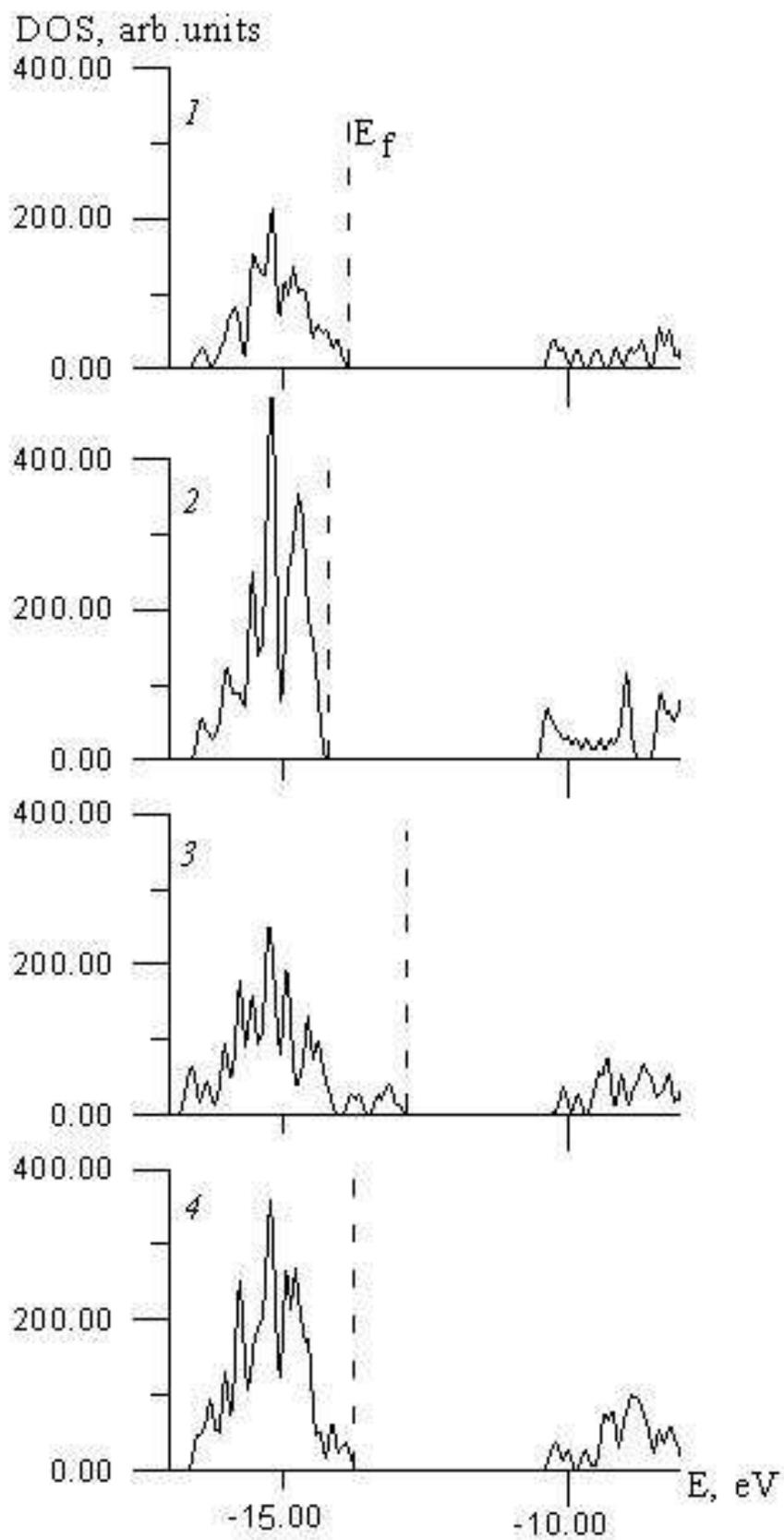}
\caption{Total DOS of {\it armchair} ({\it 8},{\it 8}), ({\it
15},{\it 15}) (1,2) and {\it zigzag} ({\it 11},{\it 0}), ({\it
15},{\it 0}) (3,4) TiO$_2$ nanotubes.}
\end{figure}

\begin{figure}
\includegraphics[width=0.65\textwidth]{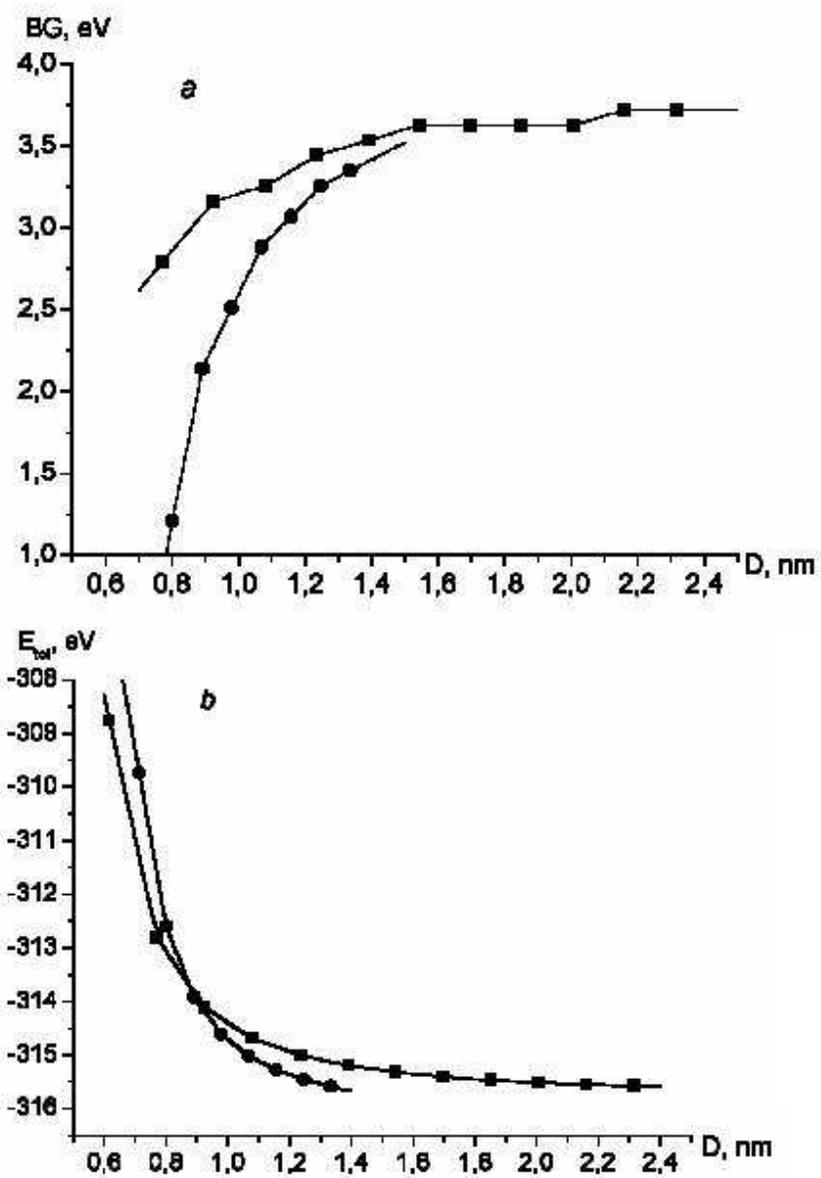}
\caption{Band gap (a) and total energies (per TiO$_2$ unit)(b) of
{\it armchair} ({it\ n},{\it n}) - squares  and {\it zigzag}
({\it n},{\it 0}) - circles as a function of the diameter of Ti
cylinder.}
\end{figure}

\end{document}